\documentclass[prl,twocolumn,superscriptaddress,showkeys,nofootinbib]{revtex4}
\usepackage{amssymb,epsf,epsfig,amsmath,color}

\begin{document}

\title{\boldmath Quark forces from hadronic spectroscopy}

\author{Dan Pirjol} 

\affiliation{National Institute for Physics and Nuclear Engineering, 
Department of Particle Physics, 077125 Bucharest, Romania}

\author{Carlos Schat}
\affiliation{CONICET and Departamento de F\'{\i}sica, FCEyN, Universidad de Buenos Aires,
Ciudad Universitaria, Pab.1, (1428) Buenos Aires, Argentina}

\date{\today}

\begin{abstract}
We consider the implications of the most general two-body quark-quark interaction 
Hamiltonian for the spin-flavor structure of the negative parity $L=1$ excited baryons. 
Assuming the most general two-body quark interaction Hamiltonian, we derive two
correlations among the masses and mixing angles of these states, which constrain the mixing 
angles, and can be used to test for the presence of three-body quark interactions.
We find that the pure gluon-exchange model is disfavored by data, independently on
any assumptions about hadronic wave functions.
\end{abstract}



\maketitle

The constituent quark model (CQM) \cite{De Rujula:1975ge,Isgur:1978xj,Isgur:1999jv,Jaffe:1976ih} 
is a popular and time-tested approach used for modeling hadron properties. The
basic assumption is that the quarks inside the hadron can be approximated as nonrelativistic 
point particles with constituent masses, interacting through two-body potentials. 

In a recent paper \cite{Pirjol:2007ed} we presented a general method for relating the
quark interaction Hamiltonian to the spin-flavor structure of the hadronic mass operator. 
Consider a given
two-body interaction Hamiltonian $V_{qq} = \sum_{i>j} O_{ij} R_{ij}$, 
where $O_{ij}$ acts only on the spin-flavor indices of the quarks $i,j$, 
and $R_{ij}$ acts only on their orbital degrees of freedom. Then the hadronic 
matrix elements of the Hamiltonian $V_{qq}$ on a baryon state
$|B\rangle$ contain only the projections ${\cal O}_\alpha$ of $O_{ij}$ onto
irreducible representations of $S_3$, the permutation group of $3$ objects acting on the
spin-flavor degrees of freedom
\begin{eqnarray}\label{Vqqeff}
\langle B | V_{qq} | B\rangle = \sum_{\alpha} C_\alpha \langle B |{\cal O}_\alpha |B\rangle \, .
\end{eqnarray}

The coefficients $C_\alpha$ are related to the reduced matrix elements of the orbital operators
$R_{ij}$, and are given by overlap integrals of the quark model 
wave functions. The relation Eq.~(\ref{Vqqeff}) allows a general study of the hadronic 
spin-flavor structure independently on the orbital structure of the interaction and wave functions.
An application of the $S_3$ group in a similar
context was discussed in Ref.~\cite{Collins:1998ny}, where  
$S_3$ refers to permutations of the quarks' orbital degrees of freedom. The 
present analysis makes crucial use of the transformation properties of operators and
states under $S_3$ acting on the spin-flavor degrees of freedom.

One of the basic assumptions of the constituent quark model is the dominance 
of 2-body quark interactions. However, 3-body quark interactions may well be
present as well. For example, it is known that in QCD with $N_f=3$
light quark flavors, 3-body interactions are induced by instanton effects
('t Hooft interaction) \cite{Schafer:1996wv}.
We point out that the system of the negative parity excited nucleons is a possible
testing ground for the presence of 3-body quark forces.

In this Letter we use the representation Eq.~(\ref{Vqqeff}) to obtain information
about the spin-flavor structure of the quark forces from the system of the negative
parity excited baryons. We derive universal correlations among masses and mixing
angles which are valid in any model for quark interactions containing only two-body
interactions. 
Deviations from these predictions can be used to test for the presence of
3-body quark interactions. We obtain constraints on the strength of spin-orbit
interactions, which can be used to distinguish between two popular
models for quark interactions: the one-gluon exchange model \cite{De Rujula:1975ge}, 
and the Goldstone boson exchange model \cite{Glozman:1995fu}. This gives 
information about the relative importance of these two interactions in generating
the effective quark forces in the low energy regime.

The most general 2-body quark interaction Hamiltonian in the constituent
quark model can be written in generic form as $V_{qq} = \sum_{i<j} V_{qq}(ij)$
with
\begin{eqnarray}
V_{qq}(ij) &=& \sum_k f_{0,k}(r_{ij}) O_{S,k}(ij) + 
f_{1,k}^a(r_{ij}) O_{V,k}^a(ij) \nonumber\\
& &\hspace{2cm} + 
f_{2,k}^{ab}(r_{ij}) O_{T,k}^{ab}(ij)\,, \label{2}
\end{eqnarray}
where $O_{S}, O_V^a, O_T^{ab}$ act on spin-flavor, and $f_k(r_{ij})$ are 
functions of $r_{ij} = |{\bf r}_i - {\bf r}_j|$. Their detailed form is unimportant
for our considerations. $a,b=1,2,3$ denote spatial indices.

\begin{table*}
\begin{tabular}{|c|c||c|c|c|}
\hline
Operator & ${\cal O}_{ij}$ & $O_S$ & $O_{MS}$ &  \\ 
\hline
Scalar & 1 &  1  & $-$ & $1$ \\
              & $t_i^a t_j^a$ & $T^2 - 3C_2(F)$ & $T^2 - 3 t_1 T_c - 3C_2(F)$ 
              & $O_1, O_1-3O_2$ \\
              & $\vec s_i \cdot \vec s_j$ & $\vec S^2 - \frac94$ & $\vec S^2 - 3\vec s_1\cdot \vec S_c - \frac94$ 
              & $O_2+2O_3, O_2-O_3$ \\
              & $\vec s_i \cdot \vec s_j t_i^a t_j^a$ &  $G^2 - \frac94 C_2(F)$ & $3g_1G_c - G^2 + \frac94 C_2(F)$ 
              & $\frac{F}{4}O_1+\frac12 O_2+O_3$, \\
 & & & & ${ O_1-(3+\frac{4}{F}) O_2+\frac{4}{F}O_3}$ \\
\hline
Vector (symm) & $\vec s_i + \vec s_j$  & $\vec L\cdot \vec S$ & $3\vec L\cdot \vec s_1 - \vec L \cdot \vec S$ 
              & $O_4+O_5, 2O_5-O_4$  \\
   & $(\vec s_i + \vec s_j) t_i^a t_j^a$ & 
               $\frac12 L^i \{G^{ia}, T^a\} - C_2(F) L^i S^i$ 
               & ${2 \frac{1-F}{F}} L^i S_c^i + L^i g_1^{ia} T_c^a + L^i t_1^a G_c^{ia}$ 
               & $O_6+O_7+{\frac{F-1}{2F}}O_4, O_6+O_7-{2\frac{F-1}{F}} O_4$ \\
Vector (anti) & $\vec s_i - \vec s_j$  & $-$  & $3\vec L\cdot \vec s_1 - \vec L \cdot \vec S$ & $2O_5-O_4$ \\
              & $(\vec s_i - \vec s_j) t_i^a t_j^a$ & $-$ & $L^i g_1^{ia} T_c^a - L^i t_1^a G_c^{ia}$ 
              & $O_6 - O_7$ \\
\hline
Tensor (symm) & $\{s_i^a , s_j^b\}$ & $L_2^{ij}\{S^i, S^j\}$ & $3L_2^{ij} \{ s_1^i, S_c^j\} - L_2^{ij} \{S^i, S^j\}$ 
              & $O_8+{4}O_9, O_8-{2}O_9$ \\
   & $\{s_i^a , s_j^b\} t_i^c t_j^c$ & 
                $L_2^{ij} \{ G^{ia}, G^{ja}\}$ & 
                $L_2^{ij} g_1^{ia} G_c^{ja} - {\frac{F-1}{4F} } L_2^{ij} \{ S_c^i, S_c^j\}$ 
              & ${\frac{F-1}{2 F}}O_8 + 4O_{10}, {\frac{F-1}{F} } O_8-4O_{10}$ \\
Tensor (anti)  & $[s_i^a , s_j^b]$ & $-$  & 0 & $-$ \\
                 & $[s_i^a , s_j^b] t_i^c t_j^c$      & $-$  & 0 & $-$ \\
\hline
\end{tabular}
\caption{The most general two-body spin-flavor quark interactions and their projections 
onto irreducible representations of $S_3$, the permutation group of three objects acting
on the spin-flavor degrees of freedom. $C_2(F)=\frac{F^2-1}{2F}$ is
the quadratic Casimir of the fundamental representation of $SU(F)$. The last 
column shows the projection of each two-body operator onto the basis of 10 
operators in Eq.~(\ref{10Ops}).}
\label{table_general}
\end{table*}

We list in Table~\ref{table_general} a complete set of spin-flavor 2-body 
operators with all possible Lorentz structures allowed by the orbital angular 
momentum $L=1$.
Columns 3 and 4 of Table~\ref{table_general} list the projections of 
the spin-flavor operators $O_{S}, O_V^a, O_T^{ab}$ onto the irreducible
representations of the $S_3$ permutation group, computed as explained in 
Ref.~\cite{Pirjol:2007ed}. 
The representation content depends on the symmetry of $O_{ij}$ under the 
permutation $[ij]$: the symmetric operators $O_{ij}$ are decomposed as 
$S+MS$, and antisymmetric $O_{ij}$ as $MS+A$. 

The symmetric $S$ projection depends only on quantities acting on the entire hadron 
$S^i, T^a, G^{ia}$, while the mixed-symmetric $MS$ operators depend on operators
acting on the core and excited quarks. 
We express them in a form commonly used in the application of the $1/N_c$ expansion 
\cite{Carlson:1998vx}, according to which their matrix elements are understood to be 
evaluated on the spin-flavor state $|\Phi(SI)\rangle$ constructed as a tensor product of an 
excited quark with a symmetric core with spin-flavor $S_c=I_c$.
The antisymmetric operators contain also an $A$ projection; its 
orbital matrix element vanishes for $N_c=3$ because of T-invariance 
\cite{Collins:1998ny,Pirjol:2007ed}, such that these operators do not contribute, 
and are not shown in Table~\ref{table_general}.

The orbital matrix elements yield factors of 
$L^i, L_2^{ij} = \frac12 \{L^i, L^j\} - \frac13\delta^{ij} L(L+1)$, which are 
the only possible structures which can carry the spatial index. 

From Table~\ref{table_general} one finds that the most general form of the
mass operator in the presence of 2-body quark interactions is a linear 
combination of 10 operators
\begin{eqnarray}\label{10Ops}
& & O_1 = T^2\,,\,\, O_2 = \vec S_c^2\,,\,\, O_3 = \vec s_1\cdot \vec S_c\,,
\,\, O_4 = \vec L\cdot \vec S_c\,,\\
& & O_5 = \vec L\cdot \vec s_1\,,\quad 
    O_6 = L^i t_1^a G_c^{ia}\,,\quad O_7 = L^i g_1^{ia} T_c^a\,,\nonumber\\
& &O_8 = L_2^{ij} \{ S_c^i, S_c^j\} \,,\,\,
O_9 = L_2^{ij} s_1^i S_c^j \,,\,\, O_{10} = L_2^{ij} g_1^{ia} G_c^{ja} \,.
\nonumber
\end{eqnarray}
This gives the most general form of the hadronic mass operator Eq.~(\ref{Vqqeff})
of the negative parity $L=1$ states allowing only 2-body quark operators.

The $L=1$ quark model states include the following SU(3) multiplets: 
two spin-1/2 octets $8_\frac12, 8'_\frac12$, two spin-3/2 octets $8_\frac32, 8'_\frac32$,
one spin-5/2 octet $8'_\frac52$, two decuplets $10_\frac12, 10_\frac32$ and two singlets
$1_\frac12, 1_\frac32$. States with the same quantum numbers mix, and we define
the relevant mixing angles in the nonstrange sector as
\begin{eqnarray}
\left\{
\begin{array}{cc}
N(1535) & = \cos\theta_{N1} N_{1/2} + \sin\theta_{N1} N'_{1/2}\\
N(1650) & = -\sin\theta_{N1} N_{1/2} + \cos\theta_{N1} N'_{1/2}\\
\end{array}
\right.
\end{eqnarray}
and analogous for the $J=3/2$ states with the replacements 
$(N(1535),N(1650),N_{1/2},N'_{1/2},
\theta_{N1}) \to (N(1520),N(1700),N_{3/2},N'_{3/2},\theta_{N3})$.

The quark model basis states $(N_{1/2},N'_{1/2})$ and $(N_{3/2},N'_{3/2})$
have quark spin $S=(1/2,3/2)$
which adds up together with the orbital angular momentum $L=1$ to give $J=1/2$ 
and $J=3/2$, respectively. The mixing angles
can be chosen to lie in the range  $(0^\circ , 180^\circ)$ by appropriate
phase redefinitions of the hadron states.

The hadronic mass operator in the quark model basis can be written as a linear
combination of the 11 coefficients $\hat M_{ij} C_j = N^*_i$,
where we represent the octets and decuplets by their nonstrange members
$N^* = (N_{\frac12}, N'_{\frac12}, N_{\frac12}-N'_{\frac12}, N_{\frac32}, N'_{\frac32}, 
N_{\frac32}-N'_{\frac32}, N'_{\frac52}, \Delta_{\frac12,\frac32}, 
\Lambda_{\frac12,\frac32})^T$. The coefficients $\hat M_{ij}$ 
can be extracted from the tables in Ref.~\cite{Carlson:1998vx}.

The rank of the matrix $\hat M_{ij}$  is
9, which implies the existence of two universal relations among the 11 hadronic 
parameters (the masses of the 9 multiplets plus the two mixing angles) which must
hold in any quark model containing only 2-body quark interactions.

\begin{figure}[t!]
\includegraphics[width=4.25cm]{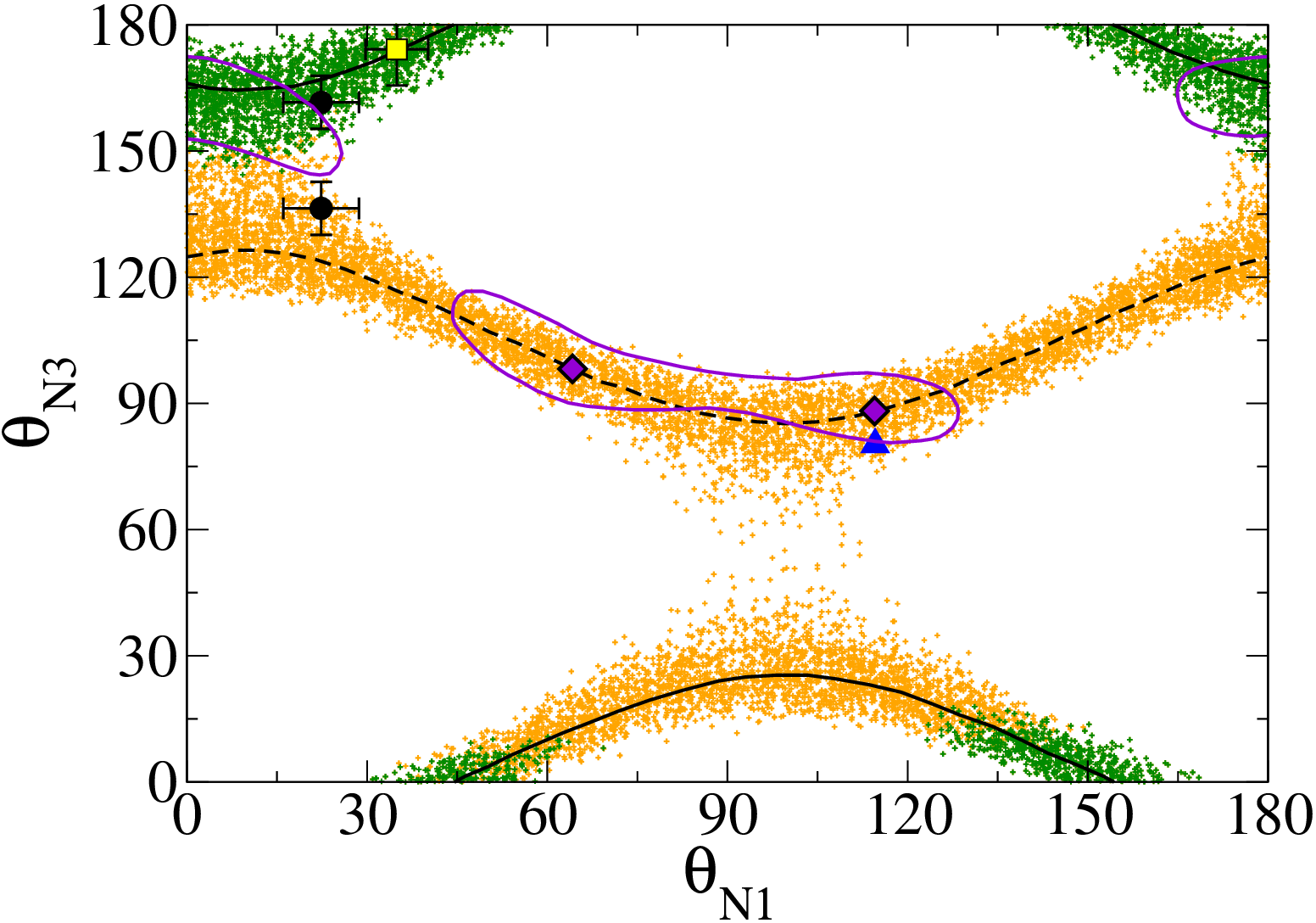}
\includegraphics[width=4.25cm]{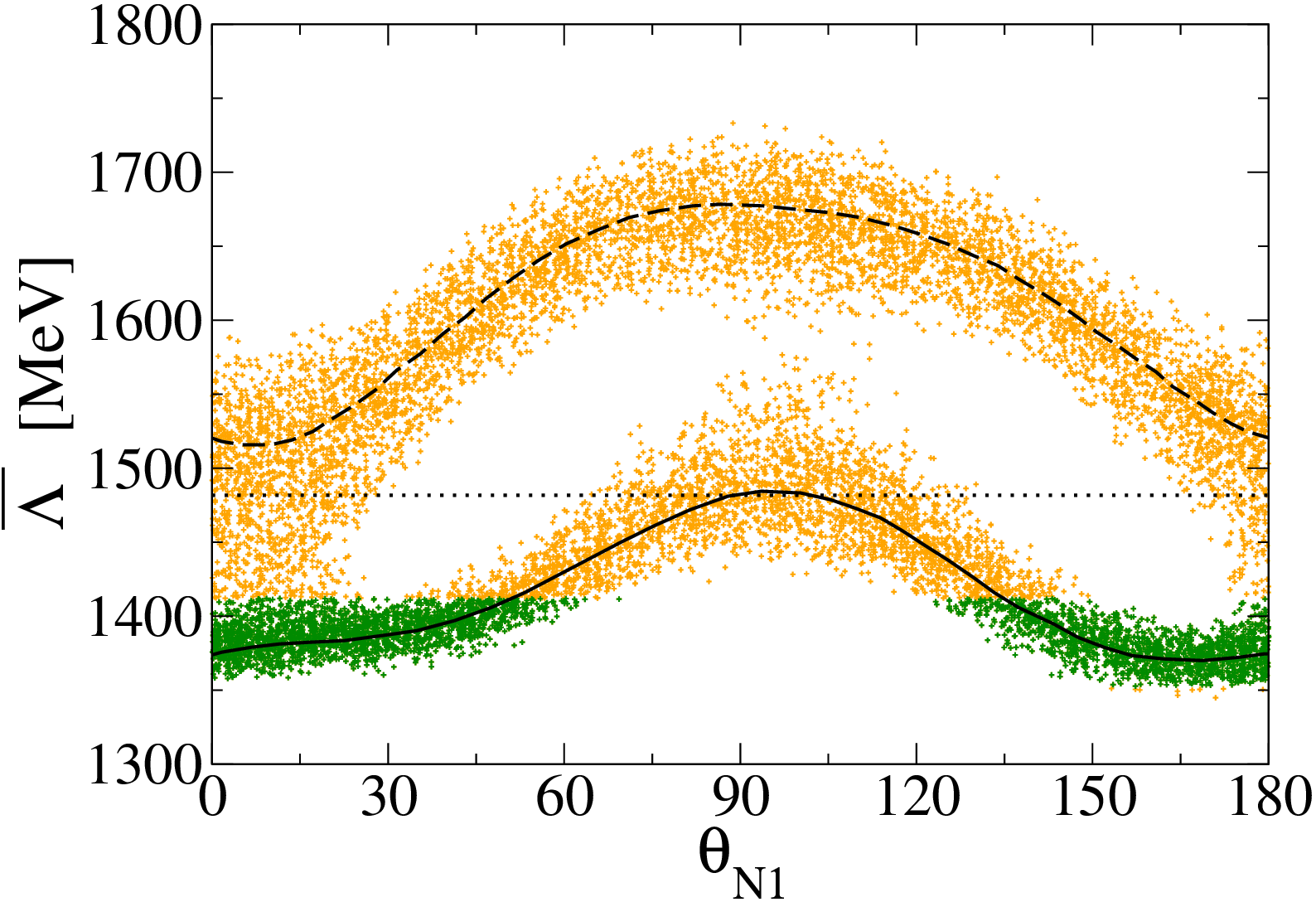}
\caption{Left: correlation in the $(\theta_{N1}, \theta_{N3})$ plane in the
quark model with the most general 2-body quark interactions. 
Right: prediction for the spin-weighted $\bar \Lambda$ mass in the 
SU(3) limit as a function of the $\theta_{N1}$ mixing angle, corresponding
to the two solutions for $\theta_{N3}$. The green points correspond to
$\bar \Lambda = \bar \Lambda_{\rm exp} - (100\pm 30)$ MeV, with
$\bar \Lambda_{\rm exp}=1481.7\pm 1.5$  MeV.}
\label{fig:OBE}
\end{figure}

The first universal relation involves only the nonstrange hadrons, and requires only 
isospin symmetry. It can be expressed as a correlation among the two mixing
angles $\theta_{N1}$ and $\theta_{N3}$ (see Fig.~\ref{fig:OBE} left)
\begin{widetext}
\begin{eqnarray}\label{OBErelation}
&& \frac{1}{2} (N(1535) + N(1650)) + \frac{1}{2}(N(1535)-N(1650))
(3 \cos 2\theta_{N1} + \sin 2\theta_{N1}) \\
&& - \frac{7}{5} (N(1520) + N(1700)) + (N(1520) - N(1700))
\Big[ - \frac{3}{5} \cos 2\theta_{N3} + \sqrt{\frac52} \sin 2\theta_{N3}\Big]
 = -2 \Delta_{1/2} + 2 \Delta_{3/2} - \frac{9}{5} N_{5/2} \nonumber\,.
\end{eqnarray}
\end{widetext}
This expresses a correlation among the mixing angles $(\theta_{N1}, \theta_{N3})$ 
which is universal for any quark model containing only 2-body 
interactions.
This correlation holds also model independently in the $1/N_c$ expansion,
up to corrections of order $1/N_c^2$, since for non-strange states the 
mass operator to order $O(1/N_c)$ \cite{Carlson:1998vx,Schat:2001xr} is generated by
 the operators in Eq.~(\ref{10Ops}).
An example of an operator which violates this correlation is $L^i g^{ja} \{ S_c^j\,,
G_c^{ia}\}$, which can be introduced by 3-body quark forces. 

On the same plot we show also the values of the mixing angles obtained in several 
analyses of the $N^*\to N\pi$ strong decays and $N^*$ hadron masses.
The two black dots correspond to the mixing angles 
$(\theta_{N1}, \theta_{N3})=(22.3^\circ,136.4^\circ)$ and
$(22.3^\circ,161.6^\circ)$ obtained from a study of the strong decays in 
Ref.~\cite{Goity:2004ss}. The second point is favored by a $1/N_c$ analysis of
photoproduction amplitudes Ref.~\cite{Scoccola:2007sn}.
The yellow square corresponds to the
values used in Ref.~\cite{Carlson:1998vx, Schat:2001xr} 
$(\theta_{N1}, \theta_{N3})=(35.0^\circ,174.2^\circ)$, 
and the triangle gives the angles corresponding to the solution $1'$ in the large $N_c$ 
analysis of Ref.~\cite{PiSc}
$(\theta_{N1}, \theta_{N3})=(114.6^\circ,80.2^\circ)$. 
All these determinations (except the triangle) are compatible with the
ranges $\theta_{N1} = 0^\circ-35^\circ, \theta_{N3} = 135^\circ-180^\circ$. 
They are also in good agreement with the  
correlation Eq.~(\ref{OBErelation}), and provide no evidence for the presence of 3-body quark 
interactions. 

The second universal relation expresses the spin-weighted SU(3) singlet mass
$\bar \Lambda = \frac16  (2\Lambda_{1/2} + 4 \Lambda_{3/2})$ 
in terms of the nonstrange hadronic parameters
\begin{widetext}
\begin{eqnarray}\label{LamAve}
\bar\Lambda &=& \frac16(N(1535) + N(1650)) + \frac{17}{15} (N(1520)+N(1700))
- \frac35 N_{5/2}(1675) - \Delta_{1/2}(1620) \\
& & - \frac16 (N(1535) - N(1650)) (\cos 2 \theta_{N1} + \sin 2\theta_{N1} )
 +
(N(1520) - N(1700)) 
(\frac{13}{15} \cos 2 \theta_{N3} - \frac13 \sqrt{\frac52} \sin 2\theta_{N3})\,.\nonumber
\end{eqnarray}
\end{widetext}
The rhs of Eq.~(\ref{LamAve}) is plotted as a function of $\theta_{N1}$ in 
the right panel of Fig.~\ref{fig:OBE}, where it can be compared against the experimental
value $\bar\Lambda = 1481.7\pm 1.5$ MeV. 
Allowing for SU(3) breaking effects $~\sim 100$ MeV, 
this constraint is also compatible with the range for $\theta_{N1}$ obtained above 
from direct determinations of the mixing angles.

Combining the Eqs.~(\ref{OBErelation}) and (\ref{LamAve}) gives a 
determination of the mixing angles from hadron masses alone, in contrast to
their usual determination from $N^*\to N\pi$ decays \cite{Hey:1974nc}.
The green area in 
 Fig.~\ref{fig:OBE} shows the allowed region for $(\theta_{N1},
\theta_{N3})$ compatible with a positive SU(3) breaking correction in 
$\bar\Lambda$ of $100\pm 30$ MeV. One notes a good agreement between this 
determination of the mixing angles and that from $N^*\to N\pi$ strong decays.

We derive next constraints on the spin-flavor structure of the quark interaction,
which can discriminate between models of effective quark interactions. There are two
popular models used in the literature, see Ref.~\cite{Isgur:1999jv} for a discussion 
in the context of the states considered here.
The first model is the one-gluon exchange model (OGE) \cite{De Rujula:1975ge}
which includes operators in Table~\ref{table_general} without isospin dependence. 
Expressed in terms of the  operator basis $O_{1-10}$ this gives the constraints
\begin{eqnarray}\label{OGEconst}
C_{1} = C_{6} = C_{7} = C_{10} = 0\,.
\end{eqnarray}

An alternative to the OGE model is the Goldstone boson exchange model (GBE) 
\cite{Glozman:1995fu}. 
In this model quark
forces are mediated by Goldstone boson exchange, and the quark Hamiltonian
contains all the operators in Table~\ref{table_general} which contain the flavor
dependent factor $t_i^a t_j^a$. The coefficients of the hadronic Hamiltonian
Eq.~(\ref{Vqqeff}) satisfy the constraints ($F=3$ is the number of light quark flavors)
\begin{eqnarray}\label{OBEconst}
C_{1} = \frac{F}{4} C_{3}\,,\quad C_{5} = C_{9} = 0\,.
\end{eqnarray}

We would like to determine the coefficients $C_i$ in the most general case,
and compare their values with the predictions of the two models 
Eqs.~(\ref{OGEconst}), (\ref{OBEconst}). However, since the rank of 
$\hat M_{ij}$ is 9, only the following combinations of coefficients can be
determined from the available data: $C_0,C_1 - C_3/2, C_2+C_3,
C_4, C_5, C_6, C_7, C_8+C_{10}/4, C_9 - 2C_{10}/3$.
In particular, as the 
coefficients of the spin-orbit interaction terms $C_{4-7}$ can be determined,
we propose to use their values to discriminate between different models of quark interaction.

They can be compared with the hierarchy expected in each model.
In the OGE model the flavor-dependent operators have zero coefficients 
$C_{6,7} \sim 0 \ll |C_{4,5}|$,  while in the GBE model the spin-orbit interaction of the 
excited quark vanishes $C_5\sim 0 \ll |C_{4,6}|$. 

The coefficient $C_5= 75.7\pm 2.7$ MeV is fixed by the 
$\Lambda_{3/2}-\Lambda_{1/2}$ splitting \cite{Schat:2001xr}. This indicates the presence of
the operators $s_i \pm s_j$ in the quark Hamiltonian, which 
is compatible with the OGE model.

\begin{figure}[t!]
\includegraphics[width=4.25cm]{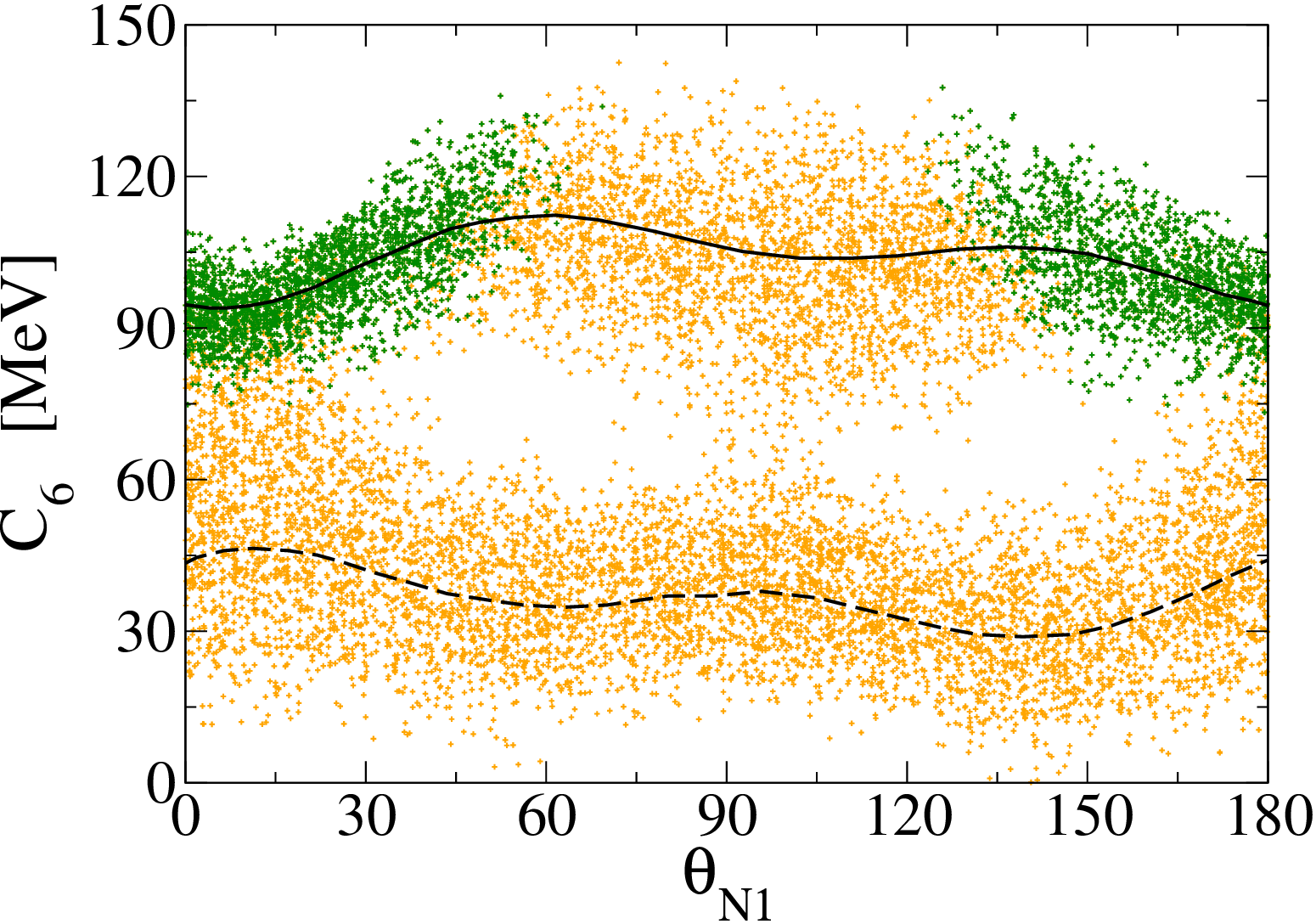}
\includegraphics[width=4.25cm]{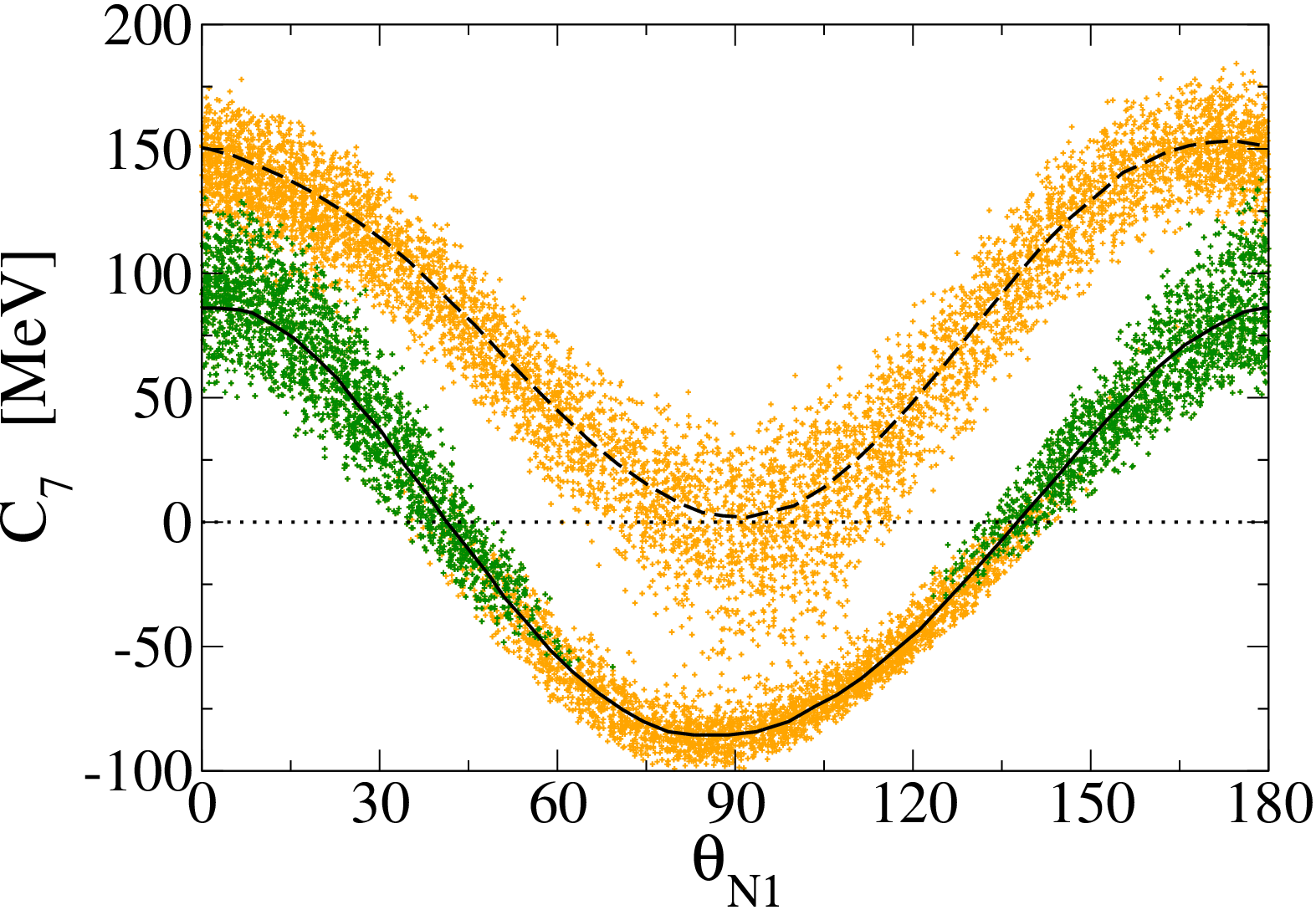}
\caption{The coefficients of the spin-orbit operators $C_{6,7}$ as functions
of the mixing angle $\theta_{N1}$, in the quark model with the most general
2-body interactions. 
The green area is obtained by
imposing the $\bar\Lambda$ constraint Eq.~(\ref{LamAve}). }
\label{fig:c4567}
\end{figure}

A suppression of the coefficients $C_{6,7}$ would be further evidence for the OGE model.
We show in Fig.~\ref{fig:c4567} the coefficients of the spin-orbit operators 
$C_{6,7}$ as functions of $\theta_{N1}$. 
Within errors small values for $C_7$ are still allowed, however no suppression
is observed for $C_6$. This indicates the presence of the operators
$(s_i \pm s_j) t^a_i t^a_j$ in the quark Hamiltonian. 
These results show that the quark Hamiltonian is a mix of the OGE and GBE
interactions. 

In the pure OGE model Eq.~(\ref{OGEconst}) the 7 nonvanishing coefficients $C_i$
can be determined from the 7 nonstrange $N^*,\Delta^*$ masses 
(assuming only isospin symmetry but no specific form of the wave functions).
This fixes the mixing angles, and the $\Lambda_{3/2}-\Lambda_{1/2}$ splitting,
up to a 2-fold ambiguity. The allowed region for mixing angles is shown as the 
violet region in Fig.~\ref{fig:OBE} left, and the central values as diamonds
$(\theta_{N1},\theta_{N3})=(64.2^\circ, 98.2^\circ), (114.5^\circ, 88.2^\circ)$.
Note that they are different from the angles obtained in 
the Isgur-Karl model $(31.7^\circ, 173.6^\circ)$ in Refs.~\cite{Isgur:1978xj,Chizma:2002qi}.

The violet region near $\theta_{N1} \sim 0$ is consistent with the
determinations from strong decays and from the SU(3) universal relation
Eq.~(\ref{LamAve}), but is ruled out by the prediction for the $\Lambda$ splitting,
in agreement with the non-zero value of $C_6$ that can be read off from
Fig.~\ref{fig:c4567}. This implies that the pure OGE model is disfavored
\footnote{Note that this argument neglects possible long-distance contributions to the 
$\Lambda$ splitting, due to the proximity of the $\Lambda(1405)$
to the $KN$ threshold. Such threshold effects are not described by the quark
Hamiltonian Eq.~(\ref{2}), and their presence could invalidate the prediction of
the $\Lambda$ splitting in the OGE model.}. 

We discussed in this Letter the predictions of the constituent quark model 
with the most general spin-flavor 2-body quark interactions, using a new
relation between the spin-flavor structure of the quark interactions and
the hadronic mass operator~\cite{Pirjol:2007ed}. 
We find two universal relations among the hadronic parameters of the negative 
parity excited baryons, valid in any model with 2-body quark interactions. 
They fix the mixing angles, and deviations from them
can probe the presence of 3-body quark interactions. We propose
new constraints on the relative importance of the different spin-flavor structures 
in the quark interaction, without imposing
any theoretical prejudice on the form of the quark interaction Hamiltonian
and the hadronic wave functions. The precision of these
constraints is limited by the uncertainty in the hadronic masses and 
mixing angles.
In principle, such information can also be gained from lattice QCD,
where the mixing angles can be related to the relative overlaps of the interpolating 
fields for the excited states.
\vspace{0.1cm}

We acknowledge useful discussions with W.~Melnitchouk and D.~Richards.

\end{document}